\documentclass[prl,english,superscriptaddress,twocolumn]{revtex4}
\usepackage{amsmath,amssymb,amsfonts,mathrsfs} 
\usepackage{amsthm}
\usepackage{CJK}
\usepackage{bm}
\usepackage{babel}
\usepackage{graphicx}

\def \K {\hat{\mathcal{K}}}
\def \Z {\mathbb{Z}}
\def \H {\mathcal{H}}

\def \T {\hat{T}}

\def \P {\hat{P}}

\def \I {\hat{I}}

\def \r {\mathbf{r}}
\def \x {\hat{\mathbf{x}}}
\def \y {\hat{\mathbf{y}}}
\def \z {\hat{\mathbf{z}}}
\RequirePackage[normalem]{ulem} 
\RequirePackage{color}\definecolor{RED}{rgb}{1,0,0}\definecolor{BLUE}{rgb}{0,0,1} 

\begin{document}

\title{$PT$ Symmetric Real Dirac Fermions and Semimetals}

\author{Y. X. Zhao}
\email[]{y.zhao@fkf.mpg.de}
\affiliation{Max-Planck-Institute for Solid State Research, D-70569 Stuttgart, Germany}
\affiliation{Department of Physics and Center of Theoretical and Computational Physics, The University of Hong Kong, Pokfulam Road, Hong Kong, China}

\author{Y. Lu}
\email[]{y.lu@fkf.mpg.de}
\affiliation{Max-Planck-Institute for Solid State Research, D-70569 Stuttgart, Germany}

\begin{abstract}
	Recently Weyl fermions have attracted increasing interest in condensed matter physics due to their rich phenomenology originated from their nontrivial monopole charges. Here we present a theory of \textit{real} Dirac points that can be understood as real monopoles in momentum space, serving as a real generalization of Weyl fermions with the reality being endowed by the $PT$ symmetry. The real counterparts of topological features of Weyl semimetals, such as Nielsen-Ninomiya no-go theorem, $2$D sub topological insulators and Fermi arcs, are studied in the $PT$ symmetric Dirac semimetals, and the underlying reality-dependent topological structures are discussed. In particular, we construct a minimal model of the real Dirac semimetals based on recently proposed cold atom experiments and quantum materials about $PT$ symmetric Dirac nodal line semimetals. 
\end{abstract}

\maketitle

\textit{Introduction--}
The discovery of topological insulators~\cite{Kane-RMP,XLQi-RMP} has galvanized the condensed matter community into research of topological phenomena in a vast variety of quantum matters. One class of materials that have attracted particular attention in recent years are the gapless topological systems, such as Weyl and various Dirac semimetals, Dirac nodal line semimetals, and nodal topological superconductors. In this new topological paradigm of gapless solid state physics, Weyl fermions are of the most fundamental status~\cite{Volovik-book,Horava,ZhaoWang-Classification}, in the sense that a Weyl point can be interpreted as a unit monopole of the $U(N)$ Berry bundle of the band structure in momentum space~\cite{Volovik-book} which, unlike other topological gapless modes, does not rely on any symmetry. The nontrivial topological charges of Weyl points as momentum space monopoles put strong constraints on the global band structure of a Weyl semimetal, leading to two primary consequences. First, a Weyl semimetal conforms the Nielsen-Ninomiya (NN) no-go theorem that Weyl points generically appear in pairs of opposite unit $U(N)$ monopole charges. This is due to the fact that the total monopole charge has to vanish due to the orientability and closeness of the first Brillouin zone (BZ) \cite{NN-NoGo,ZhaoWang-NoGo}, and a multiply charged monopole may unstably split into unit ones under perturbations~\cite{volovik-Vacuum,ZhaoWang-Classification,ZhaoWang-FermiPoint}. Second, the Chern number or TKNN invariant of a two-dimensional (2D) subsystem~\cite{TKNN,Haldane-Model} has to jump by a unit when moving across a Weyl point, and consequently the gapless chiral states originated from the Chern numbers of these 2D subsystems form Fermi arcs on the surface of a Weyl semimetal connecting the projections of Weyl points in the bulk~\cite{XGWan-WSM}, which are experimentally observable~\cite{BQLv-NP-WSM,DHong-PRX-WSM,SYXu-NP-WSM,SYXu-Science-WSM,Lu-Science-WSM}. 

In this Letter, we present a \textit{real} generalization of the Weyl semimetal through the combined $PT$ symmetry with $(PT)^2=1$, where $P$ indicates the inversion symmetry and $T$ the time-reversal. Note that $PT$ has to be broken for a Weyl semimetal unless gauge potentials are present~\cite{PT-WSM}. 
Analogous to the innumerable interesting phenomena embedded in the classification of topological insulators and superconductors~\cite{TI-Classification-I,TI-Classification-II}, the real counterpart of the Weyl point, which we call a real Dirac point, is of particular interest in the recently established classification of $PT$ and $CP$ (charge conjugate $C$) symmetric topological gapless systems~\cite{Zhao-Schnyder-Wang-PT}.
We show that the $PT$ symmetry endows a reality condition on the band structure inducing a real Berry bundle over the BZ, and the real Dirac point is actually a unit monopole for the $O(N)$ Berry bundle in contrast to the $U(N)$ bundle associated to a Weyl point. Historically, the real monopole first appeared in an entirely different context about $SO(3)$ gauge field theory in Polyakov' classic work~\cite{Polyakov-Monopole}. While the real monopole nature of the real Dirac point guarantees that all features of Weyl semimetals find their real counterparts in the real Dirac semimetal including the no-go theorem and gapless chiral Fermi arc surface states, the reality plays a nontrivial role.
In particular, the surface Fermi arcs are of richer band structure that preserves the reality.
In addition, physical systems of such $PT$ symmetry have been recently predicted in real materials
\cite{Nodal-Line-I,Nodal-Line-II,Nodal-Line-III,RuiYu-PT-DSM}
and designed in cold atom experiments
\cite{DWZ-ColdAtom-PT} for the realization of $PT$-invariant nodal line Dirac semimetals. Starting with such Dirac semimetals, we provide a general recipe for model construction of real Dirac semimetals, which may pave the way for their future experimental realization. Before moving to the detailed study, we note that throughout this work only the combined $PT$ symmetry is required, while individual $P$ and $T$ symmetries may be violated separately.

\textit{As real monopoles in momentum space--}
We first recall some basics of a Weyl point described by
\begin{equation}
	\H_W(k)=\mathbf{k}\cdot\sigma. \label{WeylPoint}
\end{equation} 
The gapless point at the origin is associated with a topologically nontrivial $U(1)$ Berry bundle on the sphere $S^2$ enclosing the gapless point, and therefore can be regarded as a monopole of $U(1)$ group. To identify the topological nature of Eq.(\ref{WeylPoint}), we analyze the Berry bundle by choosing stereographic coordinates for the north and south hemisphere, respectively, where without loss of generality we choose the unit sphere $|\mathbf{n}|=1$ (See the Supplemental Material (SM) for details~\cite{Supp}). The  Hamiltonian \eqref{WeylPoint} has eigenstates $|+,z\rangle^N=(1,z)^T$ ($|+,z\rangle^S=(\bar{z},1)^T$) and $|-,z\rangle^N =(\bar{z},-1)^T$ ($|-,z\rangle^S=(1,-z)^T$) on the north (south) hemisphere with positive and negative unit energy, respectively, where `$T$' denotes vector or matrix transposition. Negative states $|-,z\rangle^{N/S}$ on the two patches of coordinates are sections of the $U(1)$ principle fiber bundle given by the projector $P=[1-\mathrm{sgn}(\H_W|_{S^2})]/2$, which determine the transition function on the intersection $S^1$ of the two patches,
\begin{equation}
	|-,\phi\rangle^S=g_{SN}^{\mathbb{C}}(\phi)|-,\phi\rangle^N,
\end{equation}
where $g^{\mathbb{C}}_{SN}(\phi)\in U(1)$ for any point $\phi$ on the circle $S^1$.
Therefore we find
$g_{SN}(\phi)=e^{i\phi}$ with $\phi\in [0,2\pi)$,
which has the unit winding number from $S^1$ to $U(1)$. Thus the Weyl point, Eq.(\ref{WeylPoint}), corresponds to the topological nontrivial complex vector bundle. This complex vector bundle is the generator of the reduced $K$ group, $\widetilde{K}(S^2)\cong \Z$, which classifies $U(N)$ monopoles as gapless points in 3D momentum space, noting that $U(1)$ is the subgroup of $U(N)$ for any positive integer $N$. Namely the bundle for a monopole of multiple charge $n$ can be obtained from a $n$-copy direct sum of that for the Weyl point~\cite{Notes-negative-unit}.  


To give a real structure for a Berry bundle and construct a $O(N)$ monopole therein~\cite{Note-U(N)-O(N),AZee-Nonabelian-Berry-Bundle}, we need the $PT$ symmetry with $(PT)^2=1$. For a non-interacting fermionic model with (lattice) translation symmetry described by $\H(k)$, $T$ and $P$ symmetries are represented in momentum space as $\T=U_T\K\I$ and $\P=U_P\I$, respectively, where $U_{T/P}$ are unitary operators, $\K$ the complex conjugate, and $\I$ the inversion of momentum. The  combined $PT$ symmetry is then represented as $\T\P=U_{PT}\K$ with $U_{PT}=U_P U_T$, which gives a reality relation for the Hilbert space at each momentum $k$. From the viewpoint of $K$ theory, the $PT$ symmetry simply changes the classifying space of a flattened gapped Hamiltonian $\mathrm{sgn}(\H(k))$ at a specific $k$ as follows~\cite{Karoubi-book},
\begin{equation*}
	\frac{U(M+N)}{U(M)\times U(N)} \longrightarrow \frac{O(M+N)}{O(M)\times O(N)}
\end{equation*}
with $M$ ($N$) the number of conduction (valence) bands,  which symbolically illustrates the transition from complexity to reality. Furthermore, for every $k$ we can find a set of eigenstates $|\alpha,k\rangle$ for $\H(k)$, such that 
\begin{equation}
	|\alpha,k\rangle=\P\T|\alpha,k\rangle, \label{Reality}
\end{equation}
which makes a Berry bundle a real vector bundle associated to a $O(N)$ principle bundle. Note that there is no Kramers degeneracy for $(\P\T)^2=1$ in contrast to the case of $(\P\T)^2=-1$, which guarantees that the reality condition \eqref{Reality} holds for each band. 

For convenience and without loss of generality, we choose $\P\T=\K$, which simply means that the Hamiltonian is real, and a real Dirac point may be represented as
\begin{equation}
	\H_{RD}(k)=k_x\sigma_1\otimes\tau_0+k_y\sigma_2\otimes\tau_2+k_z\sigma_3\otimes\tau_0, \label{MajoranaPoint}
\end{equation}
as suggested by the Clifford algebra theory. It is noted that the real Hamiltonian density of Eq.(\ref{MajoranaPoint}) is actually that of Majorana fermion, but the spinor $\psi_D$ here is a Dirac spinor rather than a Majorana one, namely they do not agree with each other at second quantization. The topologically nontrivial real vector bundle given by the projector $P=[1-\mathrm{sgn}(\H_{RD}|_{S^2})]/2$ is actually the generator of the reduced orthogonal $K$ group, $\widetilde{KO}(S^2)\cong \Z_2$, and thus we may call the gapless point of Eq.(\ref{MajoranaPoint}) the unit real monopole in momentum space. To identify the $O(N)$ monopole charge of Eq.(\ref{MajoranaPoint}), we still adopt the stereographic coordinates for the unit sphere $S^2$ enclosing the gapless point, similar to the case of Weyl point (See the SM for details ~\cite{Supp}). Now it is a four-band theory, and the real eigenstates on the north hemisphere are $|+,1\rangle^N=(1,x,0,-y)^T$, $|+,2\rangle^N=(0,y,1,x)^T$ for positive energy and $|-,1\rangle^N=(x,-1,y,0)^T$, $|-,2\rangle^N=(-y,0,x,-1)^T$ for negative energy, while on the south hemisphere, $|+,1\rangle^S=(x,1,y,0)^T$, $|+,2\rangle^S=(-y,0,x,1)^T$, $|-,1\rangle^S=(1,-x,0,y)^T$, and $|-,2\rangle^S=(0,-y,1,-x)^T$, which satisfy Eq.\eqref{Reality}. On the intersection $S^1$ of the two hemispheres, eigenstates for negative energy are given explicitly as $|-,1\rangle^N=(\cos\phi,-1,\sin\phi,0)^T$, $|-,2\rangle^N=(-\sin\phi,0,\cos\phi,-1)^T$, $|-,1\rangle^S=(1,-\cos\phi,0,\sin\phi)^T$ and $|-,2\rangle^S=(0,-\sin\phi,1,-\cos\phi)^T$.
Then the transition function $g^{\mathbb{R}}_{NS}(\phi)\in O(2)$ with $\phi\in S^1$, which gives the relation
\begin{equation}
	|-,\alpha\rangle^S=[g^{\mathbb{R}}_{SN}]_{\alpha\beta}|-,\beta\rangle^N,
\end{equation}
is calculated as
\begin{equation}
	g^{\mathbb{R}}_{SN}(\phi)=\begin{pmatrix}
		\cos\phi & -\sin\phi\\
		\sin\phi & \cos\phi
	\end{pmatrix}.
\end{equation}
It is transparent that the transition function as a map from $S^1$ to $O(2)$ has a unit winding number, which verifies that the real Berry bundle of Eq.(\ref{MajoranaPoint}) generates $\widetilde{KO}(S^2)\cong \Z_2$.

Recall that the topological invariant for a Weyl point is given by the famous Chern number or TKNN invariant \cite{TKNN},
$\nu=\frac{1}{2\pi i}\int_{S^2} ~\mathrm{tr}\mathcal{F},$
where the Berry curvature is derived from complex valence eigenvectors on the gapped $S^2$ enclosing the Weyl point. To formulate a topological invariant for the real monopole, similarly, we may use the real Berry curvature $\mathcal{F}_R$, which is derived from the real connection $\mathcal{A}^R_{\alpha\beta}=\langle\alpha,k|\mathrm{d}|\beta,k\rangle$ with the real eigenstates $|\alpha,k\rangle$ satisfying Eq.(\ref{Reality}).  The corresponding topological invariant for two valence bands is given by
\begin{equation}
	\nu_R=-\frac{1}{4\pi}\int_{S^2}~\mathrm{tr}(I\mathcal{F}_R)\mod 2, \label{Real-Chern}
\end{equation}
where $\mathcal{F}_R$ is the curvature for the real Berry bundle and $I=-i\lambda_2$ is the generator of the $SO(2)$ group with $\lambda_2$ being the second Pauli matrix. We may call Eq.(\ref{Real-Chern}) the real Chern number, and the `$\mathrm{mod}~2$' should be understood according to the following paragraph.

For higher-dimensional real vector bundles on $S^2$, the transition functions on the equator are maps from $S^1$ to $O(N)$, which are classified by $\pi_1(O(N))\cong \Z_2$, and nontrivial transition functions correspond to nontrivial real vector bundles. A transition function from $S^1$ to $O(N)$ can always be continuously deformed to be a map from $S^1$ to $O(2)\subset O(N)$, and the parity of the winding number gives the homotopy class of the transition function, namely $\Z_2\cong \Z/2\Z$, though two-dimensional real vector bundles have a $\Z_2$ classification since $\pi_1(O(2))\cong \Z$. In this sense the Berry bundle of the Majorana Hamiltonian, Eq.(\ref{MajoranaPoint}), is a generator of the reduce orthogonal K theory, $\widetilde{KO}(S^2)\cong \Z_2$. 

\textit{No-Go theorem--}
Following almost the same topological arguments for the NN No-go theorem of Weyl semimetals \cite{NN-NoGo,ZhaoWang-NoGo}, we can show that real Dirac semimetals of the $PT$ symmetry satisfy the No-go theorem that real Dirac points always exist in pairs for a lattice model. Since a Brillouin zone as a tori is a closed orientable manifold, when one chooses an oriented $S^2$ enclosing any real monopole, the $S^2$ also encloses the rest of real monopoles with the opposite orientation, which implies the topological charge of the real monopole is equal to the inverse of the sum of topological charges for the rest, namely
\begin{equation}
	\sum_a \nu_R^{a}=0, \label{No-Go-Theorem}
\end{equation}
where $a$ labels real monopoles in a real Dirac semimetal~\cite{Supp}.
On the other hand, in contrast to Weyl points, real Dirac points of positive or negative unit charge are topologically indistinguishable as real monopoles due to the $\Z_2$ nature. This implies there is no canonical dipole pairing of real Dirac points in the whole Brillouin zone, which is different from the situation of Weyl semimetals. 
Considering copious topological features of Weyl semimetals originated from the monopole charges, particularly the surface Fermi arc, it is also intriguing to investigate the real counterpart of them for $PT$ invariant semimetals.

\textit{Minimal Semimetal models as real dipoles--}
Now we shall construct a minimal model of real Dirac semimetals, which contains a single dipole of real Dirac points in the Brillouin zone. To motivate our model, we construct it from $PT$ symmetric Dirac nodal line semimetals, which have recently been predicted in several quantum materials and cold atom experimental proposals \cite{RuiYu-PT-DSM,DWZ-ColdAtom-PT}. Let us start with the Bloch Hamiltonian in a proposed cold atom experiment for $PT$ symmetric Dirac nodal line semimetals, 
$
\H(k)=2t\sin k_y \sigma_2+(m-f(k))\sigma_3 \label{Cold-Atom-Line}
$
with
$
f(k)=\alpha_+(\cos k_x+\cos k_z)+\alpha_-\cos k_y,
$
where the Pauli matrices $\sigma_j$ operate in every two-level lattice site, and the $PT$ operator is represented as $\P\T=\sigma_3\K$ with $(\P\T)^2=1$. Note that only the algebraic properties of the $PT$ operator are needed for the $PT$ symmetric topological classification, while resultant topological features are independent of concrete representations~\cite{Note-UTransOfPT}. 
Since a real Dirac point has a four dimensional internal space, we first double the Dirac nodal line model, $\H\otimes\tau_0$, and then introduce appropriate $PT$ invariant couplings between them which collapses a pair of overlapping nodal lines into two real Dirac points. For such couplings the only $PT$ invariant gamma matrix is $\sigma_1\otimes\tau_2$, and a possible term is therefore $\sin k_x\sigma_1\otimes\tau_2$, which leads to our real Dirac semimetal model,
\begin{multline}
	\H_{RSM}(k)=2t_x\sin k_x\sigma_1\otimes\tau_2\\+2t_y\sin k_y\sigma_2\otimes\tau_0+[m-f(k)]\sigma_3\otimes\tau_0. \label{Real-Dirac-Semimetal}
\end{multline} 
See the SM~\cite{Supp} for the corresponding tight-binding model in real space. Since the real Dirac semimetal model is constructed from a cold atom system~\cite{DWZ-ColdAtom-PT}, in principle it is experimentally realizable and all parameters in Eq.(\ref{Real-Dirac-Semimetal}) are independently tunable. For concreteness, we would like to work in a parameter region where only a single pair of Dirac points separating in the $k_z$ axis with linear dispersion relations for their coarse-gained effective theories. The parameter region may be identified with the region, $t_x\ne 0$, and $0<(m-\alpha_-)/2\alpha_+<1$ and $(m+\alpha_-)/2\alpha_+>1$, where for the Dirac nodal line model only a single Dirac nodal line lies in the $k_x$-$k_z$ sub-Brillouin zone with $k_y=0$ centered at the origin.

To prepare the discussions about the topology of the real Dirac semimetal, it is helpful to have a look at the topological features of the Dirac nodal line model in this parameter region. Since the nodal circle has nontrivial $\Z_2$ topological charge \cite{Supp}, topological invariant for one-dimensional systems along $k_y$ is nontrvial either inside the circle or outside, but not both. To identify the nontrivial region, it is sufficient to check the one at the origin of the $k_x$-$k_z$ plane, which is given by
$
h(k_y)=2t\sin k_y \sigma_2+[(m-2\alpha_+)-\alpha_-\cos k_y]\sigma_3.
$
Assuming that $2t=\alpha_-$, we find the condition for nontrivial topological invariant is given by $|m-2\alpha_+|<\alpha_-$, which is always satisfied in the aforementioned parameter region. Accordingly when boundaries are open properly perpendicular to $\y$-direction, there are drum-head states filling all the inside of the nodal loop.

\begin{figure}
	\includegraphics[scale=1]{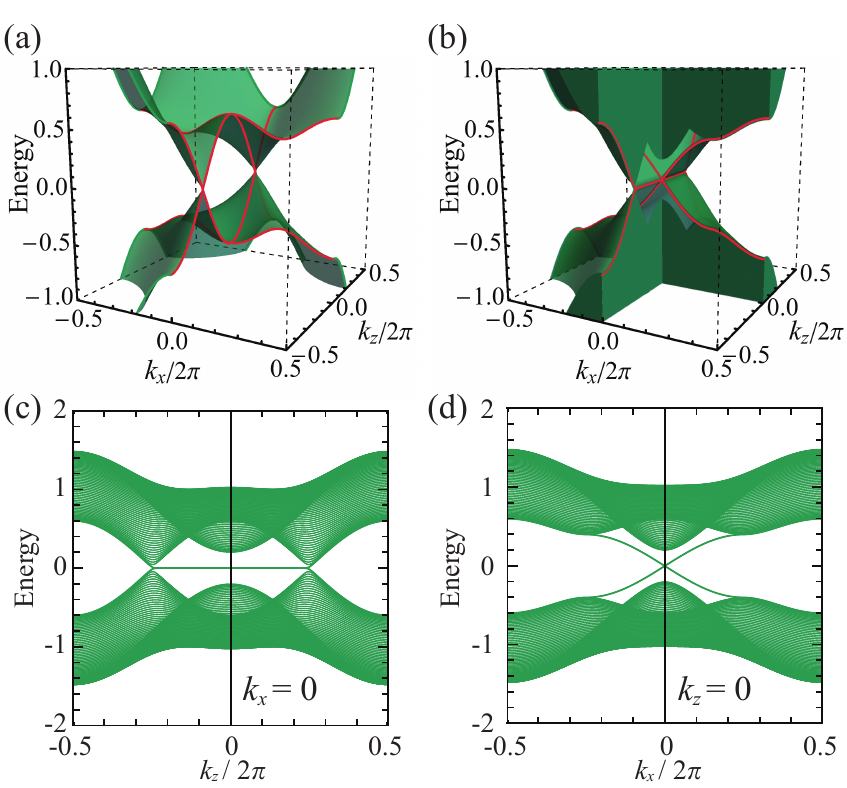}
	\caption{Spectra of the Real Dirac Semimetal. (a) Bulk spectrum with $k_y=0$. (b) Spectrum of a slab with 100 unit cells along $\y$-direction. (c) The cross section of (b) at $k_x=0$. (d) The cross section of (b) at $k_z=0$. Parameters: $\alpha_+=0.6$, $\alpha_-=0.4$, $t_x=0.2$, $t_y=0.5$ and $m=0.1$. \label{Spectra}}
\end{figure}

\textit{Real Fermi arcs on the surfaces--}
In momentum space, two-dimensional subsystems on $k_x$-$k_y$ planes away from two real Dirac points are gapped as being shown in Fig.\ref{Spectra}(a), for which we may associate the same topological invariant related to the $PT$ symmetry as for the real monopole, namely Eq.(\ref{Real-Chern}) with $S^2$ being replaced by the BZ. Since the two real Dirac points have unit topological charges, a two-dimensional subsystem has its topological invariant jumped by one when passing through a Dirac point, which means the topological invariant of subsystems with $k_z\in(k_z^-,k_z^+)$ is differentiated by one from that of subsystems with $k_z\notin[k_z^-,k_z^+]$ ($k_z^{\pm}$ are $k_z$ coordinates of two real Dirac points). For a specific $k_z$, the subsystem of our model \eqref{Real-Dirac-Semimetal} has the Hamiltonian, $\H_{k_z}^{2D}(k_x,k_y)=\H_{RSM}(k_x,k_y,k_z)$.

\begin{figure}
	\includegraphics[scale=1]{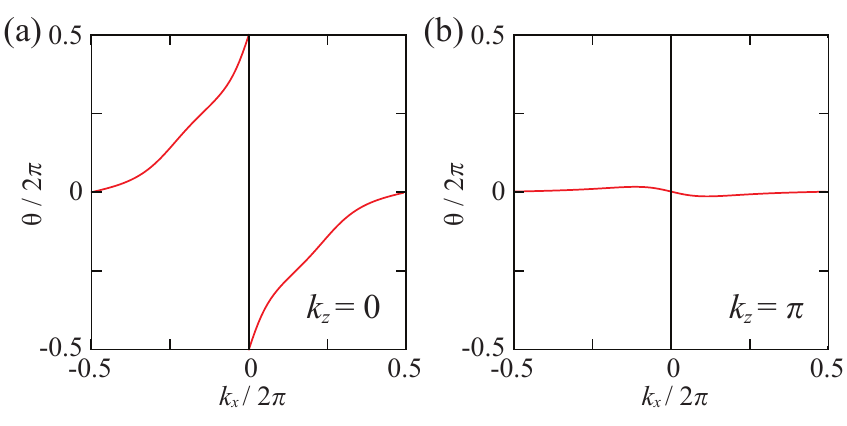}
	\caption{The windings of Wilson loops around $k_x$ for 2D subsystems with given $k_z$. Each Wilson loop is computed along a large circle parametrized by $k_y$ for fixed $k_x$ and $k_z$. Panel (a) for the 2D subsystem with $k_z=0$ has unit winding number, but (b) for $k_z=\pi$ has zero winding number.\label{Wilson-loop}}
\end{figure}

\begin{figure}
	\includegraphics[scale=1]{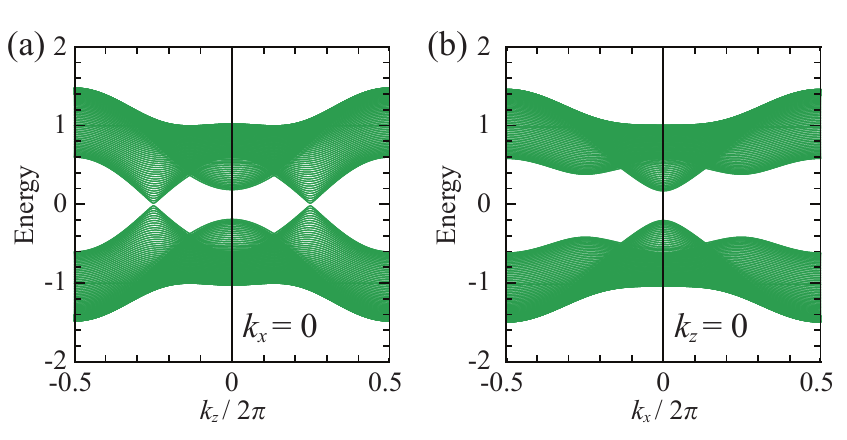}
	\caption{Spectra with $PT$ being violated. (a) and (b) are gapped after breaking the $PT$ symmetry by removing ($\uparrow A$) and ($\downarrow B$) on the top layer and ($\downarrow A$) and ($\uparrow B$) on the bottom layer, compared with (c) and (d) in Fig.1 in the main text, respectively.\label{Gapped}}
\end{figure}

Instead of computing its the topological invariant directly, we may infer it from the construction of our model. Considering the boundaries perpendicular to $\y$-direction, we start with vanishing $\sigma_1\otimes\tau_2$ term, namely $t_x=0$ in Eq.(\ref{Real-Dirac-Semimetal}), where for the bulk spectrum two nodal loops overlaps degenerately in the $k_x$-$k_z$ plane with $k_z=0$, so do two copies of drum-head states (see the previous paragraph). Then turning on the term $2t_x\sin k_x\sigma_1\otimes\tau_2$ with an infinitesimal $t_x$, degeneracies in the spectrum should be sensitive to the perturbation in contrast to the gapped regions that are insensitive. Consequently, on the surface spectrum, the region outside the drum head may be still gapped, while only a degenerate line segment survives connecting two remaining real Dirac points in the bulk under the perturbation, since it is known from above discussions that two-dimensional systems parametrized by $k_z$ are topologically trivial or nontrivial separated by two real Dirac points. Thus we infer that the Hamiltonian $\H_{k_z}^{2D}(k_x,k_y)$ is topologically nontrivial when $k_z\in(k_z^-,k_z^+)$. In fact this inferred result can be confirmed by direct calculation of the real Chern number, which is given by Eq.(\ref{Real-Chern}) with $S^2$ being replaced by the sub BZ. Alternatively, one may work out the transition function of the real bundles for 2D subsystems through the Wilson loops
\begin{equation}
	W(k_x)=P \exp \int dk_y~\mathcal{A}^R(k_x,k_y)\in O(N)
\end{equation} 
(with $P$ indicating the path order) along large circles coordinated by $k_y$, and then check the winding number around the large circle parametrized by $k_x$~\cite{Supp}. The numeric results are illustrated in Fig.\ref{Wilson-loop}.

The spectrum of Eq.(\ref{Real-Dirac-Semimetal}) with $\y$-direction being confined as a slab under natural boundary conditions respecting the $PT$ symmetry has been shown in Fig.\ref{Spectra}(b), (c) and (d). It is observed that for a surface the Fermi arc spectrum consists of two inclined planes with different angles connecting conduction and valence bands, where their crossing line links two real Dirac points in the bulk, as illustrated by Fig.\ref{Spectra}(b) and (d), in contrast to the Fermi arc of a Weyl semimetal, which contains only one inclined plane~\cite{XGWan-WSM}. As low-energy degrees of freedom, Fermi arcs on both surfaces have to preserve the reality of $PT$ as a whole, while either cannot do it alone, since $P$ maps one surface to the other and $T$ is an internal transformation. 
However, it is noted that the topological protection of the surface Fermi arcs requires that the boundary conditions have to respect the $PT$ symmetry, since the reality of the whole system has to be preserved. For instance, the gapless boundary Fermi arcs are eliminated as shown in Fig.\ref{Gapped}, after removing degrees of ($\uparrow A$) and ($\downarrow B$) on the top layer and ($\downarrow A$) and ($\uparrow B$) on the bottom layer of the slab [$\sigma_j$ and $\tau_j$ act on ($\uparrow,\downarrow$) and ($AB$), respectively], for which the $PT$ symmetry is broken, since $\P=\sigma_3\I$ maps ($\uparrow A$) [($\downarrow B$)] on the bottom (top) layer to ($\uparrow A$) [($\downarrow B$)] on the top (bottom) layer.


\textit{Summary--} The real Dirac point has been identified as the real monopole of $\Z_2$ type in the real Berry bundle with the reality being endowed by the $PT$ symmetry. The real Dirac semimetal is studied with the essential role of reality being elucidated for global topology in the whole BZ and the non-chiral surface  Fermi arcs. 
\begin{acknowledgments}
	\emph{Acknowledgments--}
	The authors thank A. P. Schnyder for discussions. 
\end{acknowledgments}

\bibliographystyle{apsrev}
\bibliography{Real-Monopole}

\begin{thebibliography}{33}
\bibitem[{Not({\natexlab{a}})}]{Notes-negative-unit}
\bibinfo{note}{The inverse of the generator can be obtained by a reflection, so
  that the direct sum of the original bundle on the surrounding $S^2$ and the
  one from reflection is a trivial bundle that serves as the zero element.}
\bibitem[{Not({\natexlab{b}})}]{Note-U(N)-O(N)}
\bibinfo{note}{For O(N) [U(N)] Berry bundles, the clutching functions of the
  corresponding principle bundles are valued in the orthogonal (unitary) group
  O(N) [U(N)], or equivalently the Berry connections are valued in the Lie
  algebra $o(N)$ [$u(N)$]~\cite{AZee-Nonabelian-Berry-Bundle}.}
\bibitem[{Not({\natexlab{c}})}]{Note-UTransOfPT}
\bibinfo{note}{Explicitly, the $PT$ operator $\P\T=\sigma_3\K$ is related to
  $\K$ by the unitary transformation given by $e^{i\sigma_3\pi/4}$}.
\end{thebibliography}

 \clearpage
 \newpage

 \appendix

 \begin{center}
 	\textbf{
 		\large{Supplemental Material}
 	}
 \end{center}
 
 \vspace{-0.2cm}

\section{The complex bundles for a Weyl point}
\begin{figure}
	\includegraphics[scale=.3]{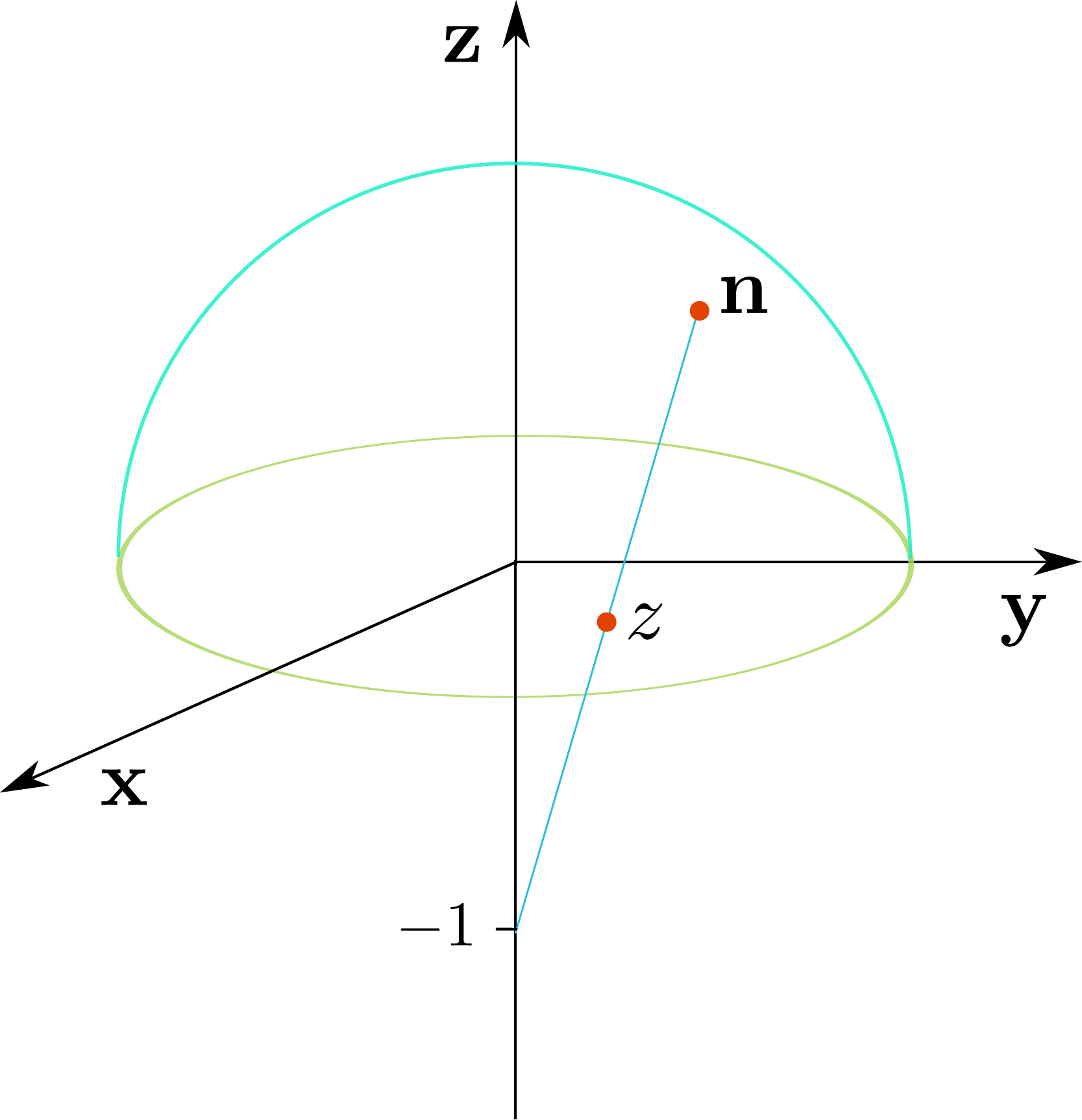}
	\caption{Stereographic coordinates of the north hemisphere. \label{stereo}}
\end{figure}

To identify the topological nature of Eq.(1) in the main text, we analyze the Berry bundle by choosing stereographic coordinates for the north and south hemisphere, respectively, as illustrated in Fig.\ref{stereo}. Then for the north hemisphere,
\begin{equation}
	n_x+in_y=\frac{2z}{1+z\bar{z}},\quad n_z=\frac{1-z\bar{z}}{1+z\bar{z}},
\end{equation}
where without loss of generality we choose the unit sphere. Accordingly Eq.(1) restricted on the north hemisphere is expressed as
\begin{equation}
	\H^N_W=\frac{1}{1+z\bar{z}}\begin{pmatrix}
		1-z\bar{z} & 2\bar{z}\\
		2z & -1+z\bar{z}
	\end{pmatrix},
\end{equation}
which has eigenvectors
\begin{eqnarray}
	|+,z\rangle^N &=& \frac{1}{\sqrt{1+z\bar{z}}}\begin{pmatrix}
		1\\ z
	\end{pmatrix},\\
	|-,z\rangle^N &=& \frac{1}{\sqrt{1+z\bar{z}}}\begin{pmatrix}
		\bar{z}\\-1
	\end{pmatrix},
\end{eqnarray}
for eigenvalues $\pm 1$, respectively. Similarly for the south hemisphere,
\begin{equation}
	n_x+in_y=\frac{2z}{1+z\bar{z}},\quad n_z=\frac{1-z\bar{z}}{1+z\bar{z}},
\end{equation}
and
\begin{equation}
	\H^S_W=\frac{1}{1+z\bar{z}}\begin{pmatrix}
		-1+z\bar{z} & 2\bar{z}\\
		2z & 1-z\bar{z}
	\end{pmatrix},
\end{equation}
with
\begin{eqnarray}
	|+,z\rangle^S &=& \frac{1}{\sqrt{1+z\bar{z}}}\begin{pmatrix}
		\bar{z}\\ 1
	\end{pmatrix},\\
	|-,z\rangle^S &=& \frac{1}{\sqrt{1+z\bar{z}}}\begin{pmatrix}
		1\\-z
	\end{pmatrix}.
\end{eqnarray}

\section{The real bundle for a Real Dirac point}
On the north hemisphere, the Hamiltonian of Eq.(4) may be represented as
\begin{eqnarray}
	\H_{RD}^{N,\uparrow}&=&\frac{1}{1+z\bar{z}}\begin{pmatrix}
		1-z\bar{z} & 2\bar{z}\\
		2z & -1+z\bar{z}
	\end{pmatrix},\\
	\H_{RD}^{N,\downarrow}&=&\frac{1}{1+z\bar{z}}\begin{pmatrix}
		1-z\bar{z} & 2z\\
		2\bar{z} & -1+z\bar{z}
	\end{pmatrix},
\end{eqnarray}
for the eigenspaces of $\tau_y=\pm 1$, respectively. Then a set of eigenstates can be worked out straightforwardly as
\begin{eqnarray}
	|+,\uparrow\rangle^N &=& \begin{pmatrix}
		1\\z
	\end{pmatrix}\otimes\begin{pmatrix}
	1\\i
\end{pmatrix},\\
|+,\downarrow\rangle^N &=& \begin{pmatrix}
	1 \\ \bar{z}
\end{pmatrix}\otimes \begin{pmatrix}
1\\-i
\end{pmatrix},
\end{eqnarray}
\begin{eqnarray}
	|-,\uparrow\rangle^N &=& \begin{pmatrix}
		\bar{z}\\-1
	\end{pmatrix}\otimes\begin{pmatrix}
	1\\i
\end{pmatrix},\\
|-,\downarrow\rangle^N &=&\begin{pmatrix}
	z \\ -1
\end{pmatrix}\otimes \begin{pmatrix}
1\\-i
\end{pmatrix}.
\end{eqnarray}
In parallel for the south hemisphere,
\begin{eqnarray}
	\H_{RD}^{S,\uparrow} &=& \frac{1}{1+z\bar{z}}\begin{pmatrix}
		-1+z\bar{z} & 2\bar{z}\\
		2z & 1-z\bar{z}
	\end{pmatrix},\\
	\H_{RD}^{S,\downarrow} &=& \frac{1}{1+z\bar{z}}\begin{pmatrix}
		-1+z\bar{z} & 2z\\
		2\bar{z} & 1-z\bar{z}
	\end{pmatrix},
\end{eqnarray}
with the eigenvectors
\begin{eqnarray}
	|+,\uparrow\rangle^S &=& \begin{pmatrix}
		\bar{z}\\1
	\end{pmatrix}\otimes\begin{pmatrix}
	1\\i
\end{pmatrix},\\
|+,\downarrow\rangle^S &=& \begin{pmatrix}
	z \\ 1
\end{pmatrix}\otimes \begin{pmatrix}
1\\-i
\end{pmatrix},
\end{eqnarray}
\begin{eqnarray}
	|-,\uparrow\rangle^S &=&\begin{pmatrix}
		1\\-z
	\end{pmatrix}\otimes\begin{pmatrix}
	1\\i
\end{pmatrix},\\
|-,\downarrow\rangle^S &=& \begin{pmatrix}
	1 \\ -\bar{z}
\end{pmatrix}\otimes \begin{pmatrix}
1\\-i
\end{pmatrix}.
\end{eqnarray}
The reality from the $PT$ symmetry is reflected by
\begin{equation}
	|E,\uparrow\rangle^{N/S}=\P\T|E,\downarrow\rangle^{N/S}.
\end{equation}
We can linearly recombine above states as sections of the real vector bundle,
\begin{eqnarray}
	|+,1\rangle^{N/S}&=& \frac{1}{2}(|-,\uparrow\rangle^{N/S}+|-,\downarrow\rangle^{N/S}),\\ |+,2\rangle^{N/S}&=&\frac{1}{2i}(|-,\uparrow\rangle^{N/S}-|-,\downarrow\rangle^{N/S}).
\end{eqnarray}

\subsection{Topology of $PT$ symmetric Dirac nodal line semimetals}
In this section, we give a brief introduction to topological structures of $PT$ symmetric Dirac nodal line semimetals, which is needed in the main text. $PT$ symmetric Fermi points or three-dimensional nodal lines are classified by the reduced orthogonal $K$ group, $\widetilde{KO}(S^1)\cong \Z_2$, which is related to the homotopy group $\pi_0(O(N))\cong \Z_2$, namely the group $O(N)$ has two components distinguished by the determinant being $1$ or $-1$. In particular $O(1)$ has only two elements $\{1,-1\}$ as a multiplicative group. A topologically nontrivial case is exemplified by
\begin{equation}
	\H_D=k_x\sigma_x+k_y\sigma_3,
\end{equation}
which may be regarded as a coarse-gained theory for a Dirac point or a local expansion toward transverse dimensions of a nodal line with $\P\T=\K$. Then on the unit circle surrounding the gapless point,
\begin{equation}
	\H_D(\phi)=\cos\phi\sigma_1+\sin\phi\sigma_3,
\end{equation}
which has the real eigenfunction for the negative energy,
\begin{equation}
	|-,\phi\rangle=\begin{pmatrix}
		\cos\frac{\phi}{2}-\sin\frac{\phi}{2}\\
		-\cos\frac{\phi}{2}-\sin\frac{\phi}{2}
	\end{pmatrix}.
\end{equation}
The eigenfunction is not continuous at $\phi=\pm\pi$, and the two points are related by the transition function $-1$, which corresponds to the nontrivial element in $\pi_0(O(1))\cong \Z_2$. Accordingly, the real vector bundle generates $\widetilde{KO}(S^1)\cong \Z_2$. An alternative way to characterize the nontrivial topology is to choose a periodic complex eigenfunction, for instance $|-,\phi\rangle^C=e^{i\phi/2}|-,\phi\rangle$, and compute the Berry phase, 
\begin{equation}
	\gamma=\oint_{S^1} d\phi~ ^C\langle-,\phi|i\partial_\phi|-,\phi\rangle^C,
\end{equation}
which is $\pi\mod 2\pi$ because of the nontrivial transition function. 

\section{A sketch of the proof of the no-go theorem}
\begin{figure}
	\includegraphics[scale=1.1]{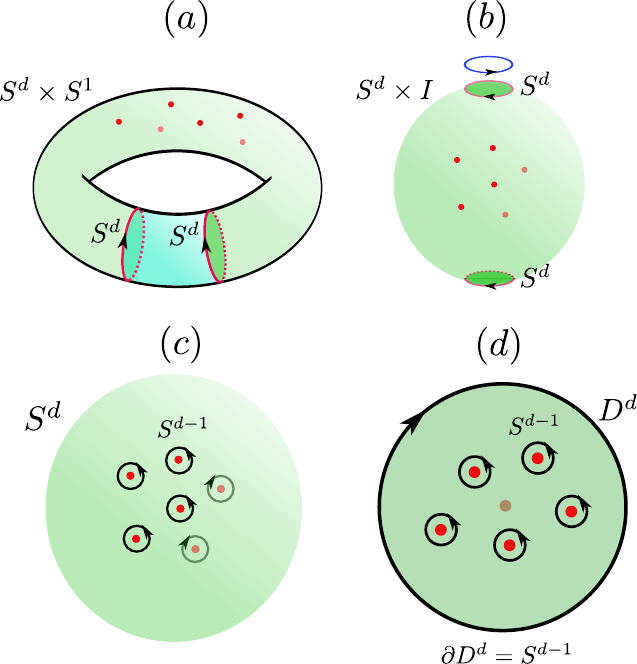}
	\caption{Deformation of torus to sphere and the vanishing of the total topological charge. \label{Torus-deformation}}
\end{figure}
In this section, we sketch the proof of the no-go theorem of the real $PT$ symmetric Dirac semimetal. We assume that in the first Brillouin zone (BZ) there are finite number of nontrivial $\Z_2$ monopoles. Through continuous deformations, the band structure can always be deformed such that only the two valence and two conduction bands that are closest to the Fermi level are curved while the others are trivial. Then the monopole charge can be calculated using the formula of the real Chern number [Eq.(7)]. If the BZ is equivalent to an $S^3$, as illustrated in Fig.\ref{Torus-deformation}(c) with $d=3$, it is obvious that the sum of topological charges of Eq.(7) is zero, similar to the vanishing of the sum of residues of a meromorphic function on the Riemann sphere. This is due to the fact that the $S^2$ surrounding an arbitrary real Dirac point can be regarded as enclosing the rest ones with an opposite orientation, but the differential form of the Chern character is inversed when the orientation of the base manifold $S^2$ is inversed.

The reason that a torus $T^d$ is equivalent to a sphere $S^d$ is illustrated in Fig.\ref{Torus-deformation}(a) and (b). Consider a manifold $ S^{d}\times S^1$ with finite number of singular points on it, whose topological charge can be defined by an integral of a differential form on an $S^{d}$. One can always cut off an interval of a cylinder as illustrated in Fig.\ref{Torus-deformation}(a), and the resultant edges consisting of two $S^d$'s have the same topological charge since they can be smoothly moved to each other. Then from the view point of Fig.\ref{Torus-deformation}(b), the two $S^d$'s have the opposite orientations, and therefore their topological charges are canceled out, which implies the $S^{d}\times S^1$ is equivalent to $S^{d+1}$. Iteratively one may infer that $T^d=S^1\times\cdots \times S^1$ is equivalent to $S^d$.

\section{The Topological Invariant from Wilson Loops}
For 2D gapped subsystems, it is convenient to formulate the topological invariant in terms of Wilson loops. Since we do not consider weak topological insulators, the 2D BZ torus may be regarded as a sphere. Similar to the case of Fermi point, there is a transition function from $S^1$ to $O(N)$ for the real Berry bundle, which captures the topology of a 2D gapped system.  The Wilson loops are given by
\begin{equation}
	W(k_x)=P \exp \int dk_y~\mathcal{A}^R(k_x,k_y)\in O(N),
\end{equation}
where the path-ordered integration is over large circles parametrized by $k_y$ for each $k_x$, and $\mathcal{A}^R$ is the real Berry connection. The transition function is just $W(k_x)$ for $k_x$ in the circle $S^1$ of $[-\pi,\pi)$.

\section{The minimal model of real Dirac semimetal in real space}
The tight-binding version $\H_{RSM}(k)$ [Eq.(9)] up tp chemical potential terms is given by
\begin{equation}
	\begin{split}
		H= & -\frac{1}{2} \alpha^+ \sum_{\r,\tau,\sigma} \zeta_\sigma (a^\dagger_{\r,\tau,\sigma}a_{\r+\x,\tau,\sigma}+a^\dagger_{\r,\tau,\sigma}a_{\r+\z,\tau,\sigma})+\mathrm{h.c.} \\
		& -\frac{1}{2} \alpha^- \sum_{\r,\tau,\sigma} \zeta_\sigma  a^\dagger_{\r,\tau,\sigma}a_{\r+\y,\tau,\sigma}+\mathrm{h.c.} \\
		&+t_x\sum_{\substack{\r,\tau,\sigma \\ \tau'\neq\tau}}(a^\dagger_{\r,\tau,\sigma}a_{\r+\x,\tau',\bar\sigma}-a^\dagger_{\r,\tau',\sigma}a_{\r+\x,\tau,\bar\sigma})+\mathrm{h.c.}\\
		&+t_y\sum_{\r,\tau,\sigma} \zeta_\sigma a^\dagger_{\r,\tau,\sigma}a_{\r+\y,\tau,\bar\sigma}+\mathrm{h.c.}\\
		&+\sum_{\r,\tau,\sigma}\zeta_\sigma m a^\dagger_{\r,\tau,\sigma}a_{\r,\tau,\sigma}
	\end{split}
\end{equation}
with $\zeta_{\uparrow/\downarrow}=\pm 1$.

\end{document}